\documentclass[epj, referee]{svjour}
\usepackage{amsmath, graphics}

\begin{document}


\title{ X-ray Resonant Magnetic Scattering : Polarisation Dependence in the non-spherical case.}

\author{Alessandro Mirone\inst{1} }
\institute{ \inst{1}European Synchrotron Radiation Facility, BP 220, F-38043 Grenoble Cedex, France\\
}

\date{Received: date / Revised version: date}

\abstract{  We develop a simple tensorial contraction  method to obtain analytical formula 
for X-ray resonant magnetic scattering. We apply the method considering  first 
electric dipole-dipole and electric quadrupole-quadrupole scattering in the isolated atom approximation 
and compare the results with previous works. 
Then we apply the method to derive phenomenological original formulas which account also 
for non-spherical systems  and for dipole-quadrupole mixing.
}
\authorrunning{A. Mirone}
\titlerunning{Resonant Magnetic Scattering  in the non-centrosymmetric case. }

\maketitle

\section{INTRODUCTION}

X-ray  magnetic  scattering  exhibits huge enhancement at resonances corresponding 
to localized electronic states\cite{laan,CarraThole}. Because of its coherent nature and  dependence on 
polarisation,  X-ray resonant  magnetic  scattering  (XRMS) can be 
used to determine both spatial  distribution and orientation of matter magnetisation.

To design experiments  and analyse data, there is a strong need for simple analytical
formula which directly relate  the experimental geometry and sample orientation to the measured
scattered intensity.

Simple formula for RMXS amplitude can be found in the literature with  the approximation of spherical atoms\cite{morrow}.

They are composed of a sum of terms, each term being the product 
of a geometrical expression, containing polarisations,  multiplied by a  complex valued  function of energy.
In a phenomenological analysys, where one is interested in separating the different contributions to the spectra, one
can take theses complex valued function as free parameters, under the constraint that dispersion relation are satisfied.

We derive 
 a simple and  understandable theoretical fra-mework within which 
we recover   previous XRMS formula and then we  extend such formula to non-spherical systems.

\section{introduction to scattering factors}

The interaction between matter and  a photon described by a wave vector ${ k}$ and polarisation ${ \epsilon}$,
  is written, discarding elastic Thomson scattering ( the $ A^2(r)$ term in the Schroedinger equation which becomes dominant off-resonance),
and discarding also the spin-magnetic field interaction~\cite{blume} ,   as :
\begin{align}
&H_{int} =  \\
& \left( \frac{2 \pi \hbar c^2 }{ \omega V}\right)^{1/2}
({  a^\dagger }_{ k,\epsilon} exp(-i {{\bf k} \cdot {\bf r}} )+a_{k ,\epsilon} exp(i { {\bf k} \cdot {\bf  r}} ))\frac{e}{c m} {  {\bf p} \cdot  {\bf \epsilon} } \label{fondamentale}
\end{align}

where ${ r}$ is the electron coordinate, ${ p}$ the kinetic moment, ${  a^\dagger }_{ k,\epsilon}$, $a_{ k ,\epsilon}$ 
are creation annihilation photon operators, $\omega= kc $, and $V$ is the space volume.

The photon-photon resonant scattering is a second order perturbative process whose resonating amplitude  per unit of time, 
for initial ${\epsilon , k }$ and final ${\epsilon^ \prime, k^\prime  }$ states, 
is obtained from  $H_{int}$  matrix elements as :
\begin{equation}
	t_{{ \epsilon  k } \rightarrow {\epsilon^ \prime k^\prime  }} =-i 
\frac{2 \pi  e^2}{ \hbar \omega m^2 V} \sum_n \frac{\left< 0 |      e^{-i { {\bf k}^\prime \cdot {\bf r}} } { {\bf  p} \cdot {\bf  \epsilon}^\prime }    | n \right>  \left< n |   e^{ i { {\bf k} \cdot {\bf r}} } { {\bf  p} \cdot  {\bf \epsilon} } | 0 \right>    }{\omega -\omega_n +i\delta}
\end{equation}

where $<n|$ is a complete set of eigenstates of matter, $<0|$  being the initial state, $\hbar \omega_n$ is the energy difference between $<n|$ and $<0|$  and we have retained only the resonating denominator. 
In the literature one always finds the scattering amplitude $F_{{ \epsilon  k } \rightarrow {  \epsilon^ \prime k^\prime  }}$which  is related
to the above equation by the following definition where one has factored out a $i c r_0 \lambda /V$  factor  ( $r_0$ is the classical radius of the electron ) :
\begin{equation}
	t_{{ \epsilon  k } \rightarrow {  \epsilon^ \prime k^\prime  }}
= i c/V r_0 \lambda F_{{ \epsilon  k } \rightarrow {  \epsilon^ \prime k^\prime  }}
\end{equation}

or

\begin{equation}
	 F_{{ \epsilon  k } \rightarrow {  \epsilon^ \prime k^\prime  }} =\frac{1}{\hbar m}
\sum_n \frac{\left< 0 |      e^{-i { {\bf  k}^\prime \cdot {\bf r}} } { {\bf  p} \cdot  {\bf \epsilon}^\prime }    | n \right>  \left< n |   e^{i {  {\bf k} \cdot {\bf r}} } { {\bf  p} \cdot  {\bf \epsilon} } | 0 \right>    }{\omega -\omega_n +i\delta} \label{scattering}
\end{equation}

 this formula is the starting point of our analysys.
To analyse such formula we expand  $ exp(ik \cdot r) p \cdot  \epsilon  $ to first order, and rewrite the electron momentum ${\bf p}$ as $\frac{i m}{\hbar} [ H,   {\bf r} ] $  :
\begin{eqnarray}
 &exp(i {  k \cdot r} ) {  p \cdot  \epsilon }
\simeq ( 1 + i    k \cdot  r  ) \frac{i m}{\hbar} [ H, {  r \cdot  \epsilon }]
 \label{expansion}   \\ 
& =
 [ H, {  r \cdot  \epsilon }] + i [ H, {  k \cdot r}  {  r \cdot  \epsilon }] /2 \nonumber
\end{eqnarray} 
where the exponential has been expanded up to first order and  $H$ is the system Hamiltonian.
Such substitution leads to :
\begin{eqnarray}
&	 F_{{ \epsilon  k } \rightarrow { \epsilon^ \prime k^\prime  }} = \nonumber \\ 
&\frac{m}{\hbar}
\sum_n \frac{\omega_n^2 \left< 0 | ({ \epsilon^\prime  \cdot r}  - \frac{i}{2} {  k^\prime  \cdot r}  {  \epsilon \cdot  r } )  | n \right>  \left< n | ( {  \epsilon  \cdot  r}  + \frac{i}{2}{  k \cdot r}  {  \epsilon \cdot  r } ) | 0 \right>    }{\omega -\omega_n+i\delta }  \nonumber \\
\end{eqnarray}

As we are going to describe atoms as spherical entities plus some kind of distortion it is interesting 
to consider the spherical group and how the angular part of the different terms of Eq.~\ref{expansion} decompose on the 
irreducible representations of such a group.
The scalar product  ${ \bf \epsilon \cdot  r} $ can be rewritten as  a sum of products of  rank 1 spherical harmonics  tensors components :
\begin{equation}
  { \bf \epsilon  \cdot  r} = \sum_{q=-1}^{q=+1}  \epsilon_q^{1*}   r_q^{1}
\end{equation}
where, given a cartesian vector ${\bf A}$, its spherical tensor components are 
\begin{eqnarray}
A_0^1 =&  A_z\nonumber\\ A_1^1 =& -(A_x+i A_y)/\sqrt(2)\nonumber\\ A_{-1}^{1} =& (A_x-i A_y)/\sqrt(2).\nonumber \\ \label{trasformazionedipolo}
\end{eqnarray}
The   product ${\bf  k \cdot r}  {\bf   \epsilon \cdot r} $  is the sum of a constant contribution ( rank 0)
and a sum  of products of rank 2 spherical harmonics tensors:
\begin{equation}
  ({  \bf k \cdot r} )({\bf  \epsilon  \cdot  r}) =  \frac{1}{3} ( {\bf  k \cdot \epsilon}) ({\bf r \cdot r})
 +  \label{tenspertens}
            \sum_{q=-2}^{q=+2} (k \epsilon)_q^{*}   (rr)_q
\end{equation}

where, given two cartesian vectors ${\bf A}$ and ${\bf B}$  , the symbols $({\bf A}{\bf B} )_q$ denote the  rank 2 spherical symmetric tensor components of their product which are :
\begin{eqnarray}
&( {\bf A} {\bf B} )_0 =&  \frac{1}{\sqrt{6}}(  2 A_z B_z    - A_x B_x  - A_y B_y     )\nonumber\\ 
&( {\bf A} {\bf B} )_{\pm 1 } =&    \mp \frac{1}{2}(   A_x B_z + A_z B_x  \pm i (  A_y B_z + A_z B_y )    )\label{trasformazionequadrupolo}\nonumber\\
&( {\bf A} {\bf B} )_{\pm 2 } =&         \frac{1}{2}(   A_x B_x - A_y B_y  \pm i (  A_x B_y + A_y B_x )    )\nonumber\\ 
\end{eqnarray}

We see from Eq.~\ref{tenspertens} that spherical tensorial components of the tensor product ${\bf k \otimes \epsilon}$ are coupled to same rank components
of the product ${\bf r \otimes r}$ to form a scalar. As the product ${\bf r \otimes r}$ has no antisymmetric component 
the rank 1 components of the product ${\bf k \otimes \epsilon}$   do not appear 
in Eq.~\ref{tenspertens}.

The transformations expressed  by Eqs. \ref{trasformazionedipolo},\ref{trasformazionequadrupolo} 
are used to simplify the scattering factor expression by choosing the angular moment  quantisation
axis in the most appropriate way.
In the next section we consider a spherical system 
pertubed by a magnetic exchange interaction  while in section \ref{esagono}
we consider a non-spherical perturbation term and derive a phenomenological expression for the magnetic scattering amplitude.

\section{derivation of  formula for the spherical case with a magnetic field perturbation }
In the case of a spherical atom perturbed by a magnetic exchange field 
the final scattering amplitude expression does not mix dipolar with quadrupolar terms
because of parity conservation. We take 
  the angular moment quantisation axis ${\bf \hat \xi}$,  along the magnetic field. 
 Terms having different quantisation number $q$ will not  mix.
The expression for scattering  then assumes the following simple form :
\begin{equation}
  F_{{ \epsilon  k } \rightarrow { \epsilon^ \prime k^\prime  }} = \sum_{q=-1}^{q=1} { F_{1,q} \epsilon_q^{1*} \epsilon_q^{1} } + \sum_{q=-2}^{q=2} { F_{2,q} (k \epsilon)_q^{*} (k \epsilon)_q}  \label{scattsumq}
\end{equation}

In the spherically symmetric case (no magnetic field) the $F_{1,q}$ and $F_{2,q}$  are independent of $q$.
The introduction of a magnetic field introduces a $q$ dependence in the scattering factors that
can be expanded as a polynomial in the  quantisation number $q$ :
\begin{eqnarray}
  F_{1,q} &=  F^{0}_1 + q F^1_1 + q^2 F^2_1 \nonumber \\
  F_{2,q} &=  F^{0}_2 + q F^1_2 + q^2 F^2_2 +  q^3 F^3_2+  q^4 F^4_2 \nonumber\\\end{eqnarray}
Where the coefficients are given by :
\begin{eqnarray}
  F^{0 \prime}_1 &=  F_{1,0} \nonumber \\
  F^{1\prime}_1 &= (F_{1,1}-F_{1,-1})/2 \nonumber \\
  F^{2\prime}_1 &=  (2  F_{1,0} -  F_{1,1} - F_{1,-1} )/2 \nonumber \\
  F^{0\prime}_2 &=   F_{2,0}\nonumber \\
  F^{1\prime}_2 &= (F_{2,-2}- F_{2,2}+  8 F_{2,1}- 8  F_{2,-1} )/12 \nonumber \\
  F^{2\prime}_2 &= ( 16 F_{2,1}+ 16 F_{2,-1}   -  F_{2,-2}  -  F_{2, 2}   - 30 F_{2,0}   )/24 \nonumber \\
  F^{3\prime}_2 &=  (   F_{2,2} -  F_{2,-2} + 2 F_{2,-1} -  2 F_{2, 1})/12 \nonumber \\
  F^{4\prime}_2 &=  (  6 F_{2,0} -  F_{2, 2} - 2 F_{2,-2} -  4 F_{2, 1}-  4 F_{2, -1})/24 \nonumber \\
\end{eqnarray}

In the non magnetic case all the terms are zero except the zero order ones.

 We can substitute $q$ in Eq.~\ref{scattsumq}
by ${\bf \hat \xi  . L}$ where ${\bf L}$ is the angular moment operator.
Going back to cartesian space, Eq.~\ref{scattsumq} is then written as 
\begin{eqnarray}
  F_{{\epsilon  k } \rightarrow {  \epsilon^ \prime k^\prime  }} = {\bf C }\left( { \bf \epsilon}^\prime  \sum_{n=0}^{n=2}{ (i {\bf \hat \xi  } \times )^n   F_1^{\prime n}} {\bf  \epsilon}     \right)+ \nonumber \\
{\bf C } \left( {\bf  k}^\prime \otimes {\bf \epsilon}^\prime  \sum_{n=0}^{n=4}{ (i {\bf\hat \xi}   \times )^n   F_2^{\prime n}} { \bf k} \otimes {\bf \epsilon}     \right)/2 \nonumber \\
\label{contraction1}
\end{eqnarray}
where the symbol $ {\bf C }$ means the sum of all possible contractions of
 the two vectors ${ \bf \epsilon^\prime}$ and ${\bf k^\prime}$ 
with the expression to their right.
 A contraction  is realised by coupling by pairs the $2N$ vectors
which enter the expression , where $2N$ is the sum of the tensors ranks. Each vectors couple is then contracted, 
the result being the scalar product,  and 
 the  $N$ scalar products are multiplied together to give the final result. The contraction of a tensor of defined rank  with
itself must not be considered. 

For dipolar scattering we have therefore 
\begin{equation}
 F_{{\bf \epsilon  k } \rightarrow { \epsilon^ \prime k^\prime  }}^{dipolar} =  { \epsilon}^\prime  \cdot  {\epsilon}  F_1^{\prime 0}  +
{ \epsilon}^\prime  \cdot ( i \hat \xi   \times { \epsilon} )  F_1^{\prime 1} +
{ \epsilon}^\prime  \cdot      (i \hat \xi   \times    i \hat \xi   \times    { \epsilon} ) F_1^{\prime 2} 
\end{equation} 
such formula  can be rearranged in the Hill \& Mc Morrow \cite{morrow} form :
\begin{equation}
 F_{{ \epsilon  k } \rightarrow {  \epsilon^ \prime k^\prime  }}^{dipolar} =  { \epsilon}^\prime  \cdot  { \epsilon}  ( F_1^{\prime 0} -  F_1^{\prime 2}) - 
 i \hat \xi  \cdot (  { \epsilon}^\prime   \times { \epsilon} )  F_1^{\prime 1} +
( { \epsilon}^\prime  \cdot   \hat \xi  )(      { \epsilon} \cdot   \hat \xi)    F_1^{\prime 2} \label{vectdipo}
\end{equation} 

For the quadrupolar term the same procedures apply with some more bookkeeping for the various terms
that arise when we apply the $i { \hat \xi } \times $ operator on the right terms. In details :
\begin{eqnarray}
  ( i { \hat \xi } \times )^0 { \epsilon } \otimes { {\bf k} } &=& { \epsilon }\otimes{ {\bf k} } \nonumber\\
  ( i { \hat \xi } \times )^1 { \epsilon }\otimes { {\bf k} } &=& i( \hat \xi \times \epsilon)  \otimes {\bf k} + i \epsilon  \otimes (\hat \xi \times {\bf k}) \nonumber\\
  -( \hat \xi \times )^2 { \epsilon } \otimes { {\bf k} }  &=& \epsilon_{\perp} \otimes  {\bf k} - 2 ( \hat \xi \times \epsilon) \otimes ( \hat \xi \times  {\bf k})
+\epsilon \otimes {\bf k}_{\perp}\nonumber \\
-i(\hat \xi \times)^3 { \epsilon } \otimes { {\bf k} } &=&i( \hat \xi \times \epsilon) \otimes  {\bf k} + 3 i \epsilon_{\perp} \otimes  (\hat \xi \times {\bf k}) \nonumber \\
 &+& 3 i  (\hat \xi  \times \epsilon ) \otimes  {\bf k}_{\perp} +i( \hat \xi \times {\bf k} ) \otimes \epsilon \nonumber\\
(\hat \xi \times)^4 { \epsilon } \otimes { {\bf k} } &=&  \epsilon_{\perp} \otimes  {\bf k} - 8(\hat \xi \times \epsilon)  \otimes  (\hat \xi \times {\bf k})  \nonumber \\ 
&+& 6 \epsilon_{\perp} \otimes  {\bf k}_{\perp} + {\bf k} _{\perp} \otimes \epsilon \nonumber \\
 \end{eqnarray}
Each of the terms of the above equation must be multiplied  by its own $F_2^{\prime q}$ and contracted with 
${  \epsilon^ \prime {\bf k}^\prime  }$. The result is :
\begin{eqnarray}
   F^{quadrupolar}_{{ \epsilon  k } \rightarrow {  \epsilon^ \prime k^\prime  }} =
( F_2^{\prime 0}+ 2 F_2^{\prime 2} +  8 F_2^{\prime 2}) S( \epsilon^{\prime} \cdot \epsilon ~ {\bf k}^\prime \cdot {\bf k}  )/4  \nonumber \\
+ i ( - F_2^{\prime 1} - 4 F_2^{\prime 3}  ) S(
  z \cdot (\epsilon^\prime \times \epsilon) ~  {\bf k}^\prime \cdot {\bf k} 
 )/2  \nonumber \\
+(- F_2^{\prime 2} - 7  F_2^{\prime 4}) S(   \epsilon^\prime \cdot z  ~   \epsilon \cdot z    ~ 
 {\bf k}^\prime \cdot {\bf k} )/2  \nonumber \\
+(- 2 F_2^{\prime 2} - 8  F_2^{\prime 4} ) S(z \cdot (\epsilon ^ \prime \times \epsilon ) ~  z \cdot ( {\bf k}^ \prime \times {\bf k} )   )/4 \nonumber \\
 +3 i  F_2^{\prime 3} S(  \epsilon^\prime  \cdot z  ~   \epsilon  \cdot z  ~  z \cdot ( {\bf k}^ \prime \times {\bf k} ) )/2 \nonumber \\
 + 6  F_2^{\prime 4}  \epsilon^\prime  \cdot z  ~  \epsilon  \cdot z  ~ {\bf k}^\prime  \cdot z  ~  {\bf k}   \cdot z \nonumber \\ 
\label{vectquad}
 \end{eqnarray}

where the expression $S(t)$ means the symmetrised expression formed by the term $t$ plus the other three derived terms 
that one obtains swapping $\epsilon^\prime$ with ${\bf k}^\prime$ and/or   $\epsilon$ with ${\bf k}$.
Our formula for quadrupolar scattering  is in agreement with  Hill \& McMorrow one\cite{morrow}.

\section{Non spherical case.}

\label{esagono}

In this section we develop a phenomenological formalism for  resonant magnetic scattering in the non-spherical case.
Our starting point is  crystal-field theory where the non sphericity of the atomic environment
is represented by a  one-particle mean field potential
which is added to the atomic Hamiltonian. Such an approach, pioneered by the works of Bequerel\cite{bequerel}, Bethe\cite{bethe}, Kramers\cite{kramers} and van Vleck\cite{vanvleck},
 was  applied to the calculation of x-ray  absorption spectra and scattering factors for the first time  by  van der Laan and Thole\cite{laan}.

The crystal field $T$ is given by a superposition of spherical tensors :
\begin{equation}
T = \sum_{l,q} t_{l,q} T^l_q
\end{equation}    

The tensor T must be invariant under all operations of the point symmetry group of the system\cite{CarraThole}.

	We treat  perturbatively the crystal-field correction to scattering amplitude
considering a  process, in one-particle approximation,
where an electron is promoted from
a closed shell state, denoted by $\left|n_gl_g\frac{1}{2};J_aj_z\right>$, to the intermediate states
  $\left|n_al_am_{az}\right>\left|\frac{1}{2}\sigma\right>$ and $\left|n_bl_bm_{bz}\right>\left|\frac{1}{2}\sigma\right>$
of two opens shells $(n_a,l_a)$ and $(n_b,l_b)$. The perturbation $T$ has matrix elements which mix
the two shells.

We write the initial one-electron state, discarding some unessential labels,  as:
\begin{equation}
  \left|J_aj_z\right> = \sum_\sigma c(j_z,\sigma) |\sigma> | l_g,j_z-\sigma> \label{core}
\end{equation}
where $c(j_z,\sigma)$ stands for the Clebsh-Gordon coefficient 
$c(l_g, j_z -\sigma, 1/2,\sigma ; J_a, j_z )$.

The electron-photon interaction is represented by the tensors $P^\prime=P(\epsilon^\prime, {\bf k}^ \prime)$ 
for the outgoing photon, and  $P$  for the incoming one.

The scattering amplitude at first order in $T$ takes the form :

\begin{align}
  \sum_{m_{gz}}  & \sum_\sigma  c^2_{(m_{gz}+\sigma,\sigma)}  \sum_{qq^\prime}   <l_g,m_{gz}|P_b^{\prime *} | l_b, m_{gz}+q^ \prime>
\times  \nonumber \\
&  f_{b(m_{gz}+q^\prime, \sigma)}   <  l_b, m_{gz}+q^\prime | T |  l_a, m_{gz}+q> 
   \times  \nonumber  \\
&  f_{a(m_{gz}+q, \sigma)} < l_a, m_{gz}+q | P_a | l_g,m_{gz}>  \nonumber  \\
+&[  a \leftrightarrow b] \label{sommalm}
\end{align}

In this expression the factor $f_b(m,\sigma)$, or $f_a(m,\sigma)$, contains implicitely
 the electron propagator for the spherical atom and accounts also for the orbital occupancies.

Starting from this expression we can arrive at a contracted form which contains, beside polarisation vectors
and the  magnetisation axis, the $T$ tensor representing the crystal field, 
 and the spherical tensors $T^{l_g}_{m_z}$ representing the core state. 
The perturbative process concerns intermediate levels of well defined angular moment $l_a$ and $l_b$.
This imposes a restrictions on the possible contractions : the transition due to a $P$ tensor from the $L=l_g$ ground state
to the $L=l_n$ excited levels implies that exactly $(l_g+rank(P)-l_n)/2$ contractions must be taken between the  tensor
and the inital ground state. 
Our one-particle approximation neglects the energy spread of the intermediate states due to many-body effect
and is therefore similar to the fast-collision approximation employed by Marri and Carra for the case 
of dipole-quadrupole scattering in a magnetoelectric crystals\cite{Marri}.

\begin{table}
\begin{tabular*}{\hsize}{c|ccc}
\hline
$c^2(m_z+\sigma,\sigma)$ for $L_3$             &   $m_z=1$  & $m_z=0$     &   $m_z=-1$         \\
\hline
 $\sigma=1/2$ &                    1          &    2/3        &    1/3                                \\
 $\sigma=-1/2$ &                 1/3          &    2/3        &    1                 \\
\hline
\end{tabular*}
\vspace{10pt}
\caption{\label{L3} values of $c(j_z,\sigma)$ at the $L_3$ resonance}
\end{table}

 We develop $f_a(q,\sigma)$ and $f_b(q,\sigma)$
in powers of $L_z$, as  in the previous section :
\begin{equation}
  f(q,\sigma) = \sum_{n} f_{n,\sigma} q^n \rightarrow \sum_{n} f_{n,\sigma} L_z^n \label{developp}
\end{equation}
The scattering amplitude is then written as :
\begin{eqnarray}
  \sum_{m_{gz},\sigma,n_a, n_b}& f^{b}_{n_b,\sigma} f^{a}_{n_a,\sigma}  c^2_{(m_{gz}+\sigma,\sigma)}\nonumber \\
{\bf C }( T^{l_g*}_{m_z} P_b^{\prime *}  |^{\leftarrow} 
   &(i \xi_m \times )^{n_b} T^{l_g}_{m_z} (i \xi_m \times )^{n_a} P_a |^{\rightarrow}  T^{l_g}_{m_z}  )\nonumber \\
+ symm.
\label{contraction2}
\end{eqnarray}


where the expression is symmetrised by the following substitutions:
\begin{equation}
symm. =  [a \leftrightarrow b]\nonumber \\
\end{equation}
and  the symbol $|^{\leftarrow}$ ($|^{\rightarrow}$) following a polarisation tensor means that the number of 
contractions of such tensor with the preceding (following) ground level tensorial object is constrained as 
discussed above. The $(i \xi_m \times )$ operator operates on all the object at its right
( remember that $(i \xi_m \times ) A ~ B  = ( (i \xi_m \times ) A)  ~ B + A ~( (i \xi_m \times ) B)$).

A formal derivation of these contraction rules is given in appendix A.

As an example we give a specialised expression for the $L_3$ edge.
We give in table \ref{L3} the coefficients $c^2(j_z,\sigma)$. 
It is useful to notice that for a given $m_z=j_z-\sigma$ the sum of two coefficients of opposite spin is constant and equal to $4/3$,
 while
the difference goes from $-2/3$ to $2/3$ with a linear dependence on $m_z$. 
This behaviour does not depend on the particular  edge that we have considered.
In general, for an edge $e$, we can always write :
$
 c^2(m_z+\sigma,\sigma) = l_e + 2 s_e ~ \sigma  m_z
$ 
where $l_e$ and $s_e$ are constants which depend on the edge. For example, at $L_3$, we have $l_e=2/3$ and  $s_e=1/3$,
while at $L_2$  $l_e=1/3$ and  $s_e=-1/6$.

It is then useful to rewrite equation \ref{contraction2} as :

\begin{eqnarray}
l_e \sum_{n_a, n_b}&   \frac{2}{3} ( f^{b}_{n_b,1/2} f^{a}_{n_a,1/2}+ f^{b}_{n_b,-1/2} f^{a}_{n_a,-1/2} ) \times \nonumber \\
\sum_{m_{gz}} {\bf C }( T^{l_g*}_{m_z} P_b^{\prime *}  |^{\leftarrow} 
   &(i \xi_m \times )^{n_b} T^{l_g}_{m_z} (i \xi_m \times )^{n_a} P_a |^{\rightarrow}  T^{l_g}_{m_z}  )\nonumber \\
+s_e \sum_{n_a, n_b}&   \frac{1}{3} ( f^{b}_{n_b,1/2} f^{a}_{n_a,1/2}- f^{b}_{n_b,-1/2} f^{a}_{n_a,-1/2} ) \times \nonumber \\
\sum_{m_{gz}} {\bf C }( T^{l_g*}_{m_z} P_b^{\prime *}  |^{\leftarrow} 
   &(i \xi_m \times )^{n_b} T^{l_g}_{m_z} (i \xi_m \times )^{n_a} P_a |^{\rightarrow}  (i \xi_m \times ) T^{l_g}_{m_z}  )\nonumber \\
+ symm.
\label{contraction3}
\end{eqnarray}
In the second term we have converted the $m_z$ quantization number into the operator $(i \xi_m \times )$ at the left of
the core-hole orbitals.
Once contracted, the above formula gives  in the general case a complicated expression.

However for terms which are zero and first order in  $(i \xi_m \times )$, 
the sum over core-hole orbitals can be simplified in an elegant way.   

The zero order terms describe anysotropy of the crystal field. The first order one 
describe crystal field induced corrections to the magnetic scattering.

 Concerning  the zero order terms, we notice that such sum looks like a trace.
 In other words if, in a given contraction
diagram, a vector $A$ is contracted with the {\it ket} ground orbital, and another vector $B$ is contracted with 
the same {\it bra} core orbital the result is equivalent
to a contraction of $A$ with $B$ because 
 the sum runs over a complete basis of an irreducible supbspace.

Concerning the first order terms,  $(i \xi_m \times )$  operates either on the core-hole orbital,
or on the intermediate tensors ( $P$, $P^{\prime}$, and crystal field tensor $T$).

The terms where $(i \xi_m \times )$ operates on the intermediate tensors can be simplified as in the zero order case
described above.

Now we consider the terms where one $(i \xi_m \times )$ operates on the core-hole orbital. 
The contraction diagrams arising from these  terms can be split in two classes. 

In the first class  $(i \xi_m \times )$ operates on a core-hole vector and the result is contracted
with another  core-hole vector. This class gives zero contribution because $(i \xi_m \times )$
is an antisymmetric operator.

The second class is given by the remaining terms which contain the factor $ {\bf v} \cdot ((i \xi_m \times ) {\bf h})$,
where ${\bf v}$ is a vector which enters the composition of  an intermediate tensors and ${\bf h}$ 
enters the composition of the core-hole tensor.  This factor can be rewritten as  $-((i \xi_m \times ) {\bf v})  \cdot {\bf h}$
and once again the sum over the core-hole orbitals disappear from the final expression.

The general expression for scattering amplitude, up to first order in $(i \xi_m \times )$ is therefore :

\begin{eqnarray}
\sum_{n_a, n_b=0}^{n_a+n_b=1 }&   a_{n_a,n_b} \times \nonumber \\
 {\bf C }( P_b^{\prime *}  
   &(i \xi_m \times )^{n_b} T^{l_g}_{m_z} (i \xi_m \times )^{n_a} P_a   )\nonumber \\
+ symm.
\label{contraction4}
\end{eqnarray}

where the $a_{n_a,n_b}$ depends linearly on the $f$ electron propagators and the exact linearity coefficients 
can be found working out the contraction diagrams.

For higher order terms this simplification is no more appliable in the same easy way. 
One could still write contractions where the core-hole tensor disappear but the formula would be complicate by the fact
that when one rewrites
$ {\bf v} \cdot ((i \xi_m \times )^2 {\bf h})$ as  $((i \xi_m \times )^2 {\bf v})  \cdot {\bf h} $, the
$(i \xi_m \times )^2$ operator cannot be factored out of the tensor composed by $v$ because
\begin{equation}
(i \xi_m \times )^2 ( {\bf v_1} ~ { \bf v_2} ) \neq      ((i \xi_m \times )^2  {\bf v_1} )~{ \bf v_2} +   {\bf v_1} ~  ((i \xi_m \times )^2{ \bf v_2})
\end{equation}

The core orbital disappears from the final expression because, in our approximation,
 our $L_3$ core state in Eq.~\ref{core} is not coupled to the valence orbitals.
Such an approximation is valid for the hard x-rays domain where the core-hole is deep.

For the sake of simplicity, in the following,
  examples  we limit ourself to consider only zero and first order contributions.

\subsection{Application :Magnetic diffraction amplitude for spiral antiferromagnetic holmium}

 We consider the case
of holmium. The
holmium crystal has a hcp structure where the atoms are embedded in a   local $D_{3h}$ symmetry environment.

 Taking a cartesian $x,y,z$ frame with $x$ along the hcp $a$ axis :
\begin{equation}
T =  t_2 (3 z^2 -(x^2 + y^2  +z^2) )  \pm t_3 (x^3 - 3 x y^2)+ .....\label{holmium}
\end{equation}
where the first omitted term of the serie is a rank 4 component. The  $\pm$ signs alternate from one $ab$ plane to the other.

The contribution of the  $ (2 z^2 -  x^2- y^2)$ , is centrosymmetric. So we consider separately
the dipole-dipole  and the quadrupole-quadrupole scattering.
Within the framework of our simplifying assumption (equation \ref{contraction4})
we find that the dipole-dipole scattering correction is proportional to 
\begin{equation}
 3 \epsilon^\prime \cdot {\bf z} ~ (i \hat{\xi_m} \times \epsilon)\cdot {\bf z} -  \epsilon^\prime \cdot (i \hat{\xi_m} \times \epsilon) -[\epsilon^\prime  \leftrightarrow \epsilon] \label{dipdipani}
\end{equation}
The second term of this expression merges 
 with the previously found form  for the scattering in the spherical approximation.
But the first term contains two scalar products with the ${\bf z}$ axis
and this adds complexity to the amplitude dependence on experimental geometry.

The quadrupole-quadrupole scattering,
discarding those terms that can be  merged with the spherical formulas, has the form :
\begin{align}
\{\{\{	\{ &  \nonumber  \\
& ({\bf k}^\prime \cdot \hat z)({\bf k} \cdot \hat z)  \epsilon^\prime \cdot (i \hat{\xi_m} \times \epsilon)  \nonumber  \\
 & +
[  i \hat{\xi} \times \epsilon  \leftrightarrow  {\bf k} ]~~\}  +[{\bf k} \leftrightarrow \epsilon]~~\}+
[[{\bf k}^\prime  \leftrightarrow \epsilon^\prime]]~~\}   \nonumber \\
& - [{\bf k},\epsilon \leftrightarrow {\bf k}^\prime, \epsilon^\prime]~~\}  \nonumber  \\
\label{quadquadani}
\end{align}
this  term also contains two scalar products with the ${\bf z}$ axis
 adding complexity to the amplitude dependence on experimental geometry.

These terms contribute, in our system, to the 
 amplitude of the  $2 n \pm q $ Bragg orders diffraction peaks, $n$ and q
being the antiferromagnetic wavevectors.
In an experiment which measures several  $2 n  \pm  q $ Bragg order diffraction peaks, the incidence angle
may vary considerably from one order to the others, and therefore we predict that the inclusion
of our correction could   improve  the fit.

The scattering factor contribution from the alternating term $\pm (x^2 - 3 x y^2)$
gives diffraction peaks at $2 n +1 + m q $ orders, $n$ and $m$ being integers and q
being the antiferromagnetic wavevector.
To simplify the following treatement, and  to give an example 
of a term contributing to the $2 n +1 +  q$ scattering,
we consider  magnetisation in the  $xy$ plane and 
we limit here to the case where, in equation \ref{contraction4},
only the $n_a=1$, $n_b=0$ term is considered.
We apply the contraction rules to formula  \ref{contraction4}     where  $P^\prime$ and $P$
are $\epsilon$ and ${\bf k} \otimes \epsilon$ tensors respectively.  
 In this we get case a scattering amplitude proportional to :
\begin{align}
   (( \nonumber  \\
	- 6i (\epsilon^\prime  \cdot \hat x) (\epsilon \cdot \hat x)    \hat \xi_m \cdot (\hat x \times k  ) \nonumber  \\
+ 6i (\epsilon^\prime  \cdot \hat x) (\epsilon \cdot \hat y)    \hat \xi_m \cdot (\hat y \times k  )  \nonumber  \\
+ 6i (\epsilon^\prime  \cdot \hat y) (\epsilon \cdot \hat x)    \hat \xi_m \cdot (\hat y \times k  ) \nonumber  \\
+ 6i (\epsilon^\prime  \cdot \hat y) (\epsilon \cdot \hat y)    \hat \xi_m \cdot (\hat x \times k  ) \nonumber  \\
) + [\epsilon \leftrightarrow k] \nonumber \\
) + [k\leftrightarrow k^\prime, \epsilon \leftrightarrow \epsilon^\prime] \nonumber  \\
\end{align}

\section{Conclusions}
 We have established a contraction method by which we obtain  phenomenological analytical
expressions for the scattering factors in terms of scalar and vectors products of the polarisation
vectors, of  the magnetisation axis and of the vectors defining the crystal field tensor. Using a perturbative approach 
 we have been able to go beyond the usual $SO2$ approximation \cite{morrow}
 considering simultaneaously the  magnetisation and  a general crystal field
tensor. 
Our method is a viable diagrammatic technique based on contraction between 
vectors.  With our method 
useful formula can be straightforwardly obtained
in terms of the polarisation vectors, the crystal field  vectors
and  the magnetisation direction.

We have applied the method to the  case of  holmium, giving the crystal field  correction for the 
quadrupole-quadru-pole scattering  and predicting  dipole-quadrupole peaks occurring at $2n+1+mq$ Bragg orders.
Our work completes  the Hill \& McMorrow  one\cite{morrow}.

\section{Acknowledgement}

I am indebted  to Sergio di Matteo for 
fruitful  and interesting discussions, and for  sharing with me his  theoretical views
and his experience.
 I am  grateful  to my collegues Laurence Bouchenoire
and Simon Brown of the X-mas beamline at ESRF  for motivating this 
paper with their research work. I thank Efim Kats,  from
the ILL theory group, and Francois de Bergevin, for the helpful critical reading of the manuscript.

\section{Appendix A: derivation of contraction rules }

We show in this section how an angular integration of a polynomial over a unit sphere surface can be expressed as a sum of  contraction diagrams.

The polinomial integration that we consider is :
\begin{equation}
	\int_S   ( \prod_i^{2 N} {\bf a}_i \cdot {\bf  n}  ) d n^2
\end{equation}
where $2 N$ is the polynomial order, ${\bf a}_i$ are vectors, and ${bf n}$ is a unit vector.
By multiplying and diving this expression by $\int exp(-r^2)  r^{2 N + 2 }  d r$ we transform it into an integral over the whole space:
\begin{align}
	\int_S  & ( \prod_i^{2 N} {\bf a}_i \cdot {\bf  n} )  d n^2 = \nonumber \\
	\int   & ( \prod_i^{2 N} {\bf a}_i \cdot {\bf  r}  ) exp(-r^2)  d {\bf r}^3  / (\Gamma(N+3/2)/2) = \nonumber \\
  \prod_i^{2 N} & {\bf a}_i \cdot {\bf  \partial_v }  (	\int   exp(-r^2+ {\bf v \cdot r})  d {\bf r}^3 ) |_{v=0} / (\Gamma(N+3/2)/2) = \nonumber \\
  \prod_i^{2 N} & {\bf a}_i \cdot {\bf  \partial_v }   ~ exp({\bf v}^2/4) |_{v=0}  \Gamma( 3/2) / \Gamma(N+3/2) \nonumber  \\
\end{align}
In this expression, when $v=0$ only the terms fully contracted remain because 
 if $ {\bf a}_l \cdot {\bf  \partial_v } $ takes down a ${\bf v}/2$ factor from the exponent, another  
$ {\bf a}_m \cdot {\bf  \partial_v } $ must be used to derive this ${\bf v}/2$ factor which otherwise gives zero contribution,
and this gives the scalar product ${\bf a}_l \cdot {\bf a}_m$.

The intermediate sum 
which appear in  equation \ref{sommalm}, can be considered as a projector $p_{l_a}$ over a tensorial space of definite rank $l_a$.
The projection over definite rank space is obtained through the $|^{\leftarrow}$ and  $|^{\rightarrow}$ constraint.
When applying contraction rules to expressions formed by tensor of a defined rank, one must bear in mind that, by definition, a defined
rank tensor gives zero when contracted with itself.

\end{document}